\documentclass [12pt,twoside]{article}           
\usepackage[left,modulo]{lineno}
\usepackage{setspace}

\countdef\arxiv=201

\ifnum\arxiv=0 \input cpreb.tex \fi

\ifnum\arxiv=1

\usepackage{epsfig,times,lscape}
\usepackage[usenames]{color}

    \ifnum\count102=1

    \fi

    \ifnum\count102=2
\fi

\newcommand{\nc}{\newcommand}

     \ifnum\count101=1
\nc{\qI}[1]{\section{{#1}}}
\nc{\qA}[1]{\subsection{{#1}}}
\nc{\qun}[1]{\subsubsection{{#1}}}
\nc{\qa}[1]{\paragraph{{#1}}}

\def\qpar{\vskip 2mm plus 0.2mm minus 0.2mm}
\def\qL{\hfill \break}
     \fi 

      \ifnum\count101=2
 \nc{\qI}[1]{\parindent=0mm \vskip 8mm 
{\centerline{\LARGE \color{red}#1}}\vskip 3mm}
\nc{\qA}[1]{\vskip 2.5mm \noindent 
{{\bf\large\color{blue}  #1}} \vskip 1mm \parindent=0mm}
 \nc{\qun}[1]{\vskip 1mm \noindent {\sl #1 }\quad }

\def\qL{\hfill \break}
\def\qpar{\vskip 2mm plus 0.2mm minus 0.2mm}

      \fi

\def\qth{\vrule height 12pt depth 0pt width 0pt}
\def\qtb{\vrule height 0pt depth 5pt width 0pt}

\nc{\qfoot}[1]{\footnote{{#1}}}

      \ifnum\count101=1
\def\qbu{\hfill \par \hskip 6mm $ \bullet $ \hskip 2mm}
\def\qee#1{\hfill \par \hskip 6mm (#1) \hskip 2 mm}
      \fi
      \ifnum\count101=2
\def\qbu{\hfill \par \hskip 4mm $ \bullet $ \hskip 2mm}
\def\qee#1{\hfill \par \hskip 4mm (#1) \hskip 2 mm}
      \fi

\def\qparr{ \vskip 1.0mm plus 0.2mm minus 0.2mm \hangindent=10mm
\hangafter=1}

     \ifnum\count101=1 
 
     \fi
     \ifnum\count101=2

  \def\qcitb#1{\noindent \hbox to 102mm{\hfill \small #1} \vskip 1mm}
      \fi

 \def\qpages#1{\count102=0{\loop\advance\count102 by 1
 \null \vfill\eject \ifnum\count102<#1 \repeat}}

\def\qth{\vrule height 12pt depth 0pt width 0pt}
\def\qtb{\vrule height 0pt depth 5pt width 0pt}

\def\qv{\vskip 0.1mm plus 0.05mm minus 0.05mm}
\def\qhu{\hskip 0.6mm}
\def\qhv{\hskip 3mm}

\def\qhw{\hskip 1.5mm}
\def\qleg#1#2#3{\noindent {\bf \small #1\qhw}{\small #2\qhw}{\it \small #3}\qv }
\newcommand{\promille}{%
  \relax\ifmmode\promillezeichen
        \else\leavevmode\(\mathsurround=0pt\promillezeichen\)\fi}
\newcommand{\promillezeichen}{%
  \kern-.05em%
  \raise.5ex\hbox{\the\scriptfont0 0}%
  \kern-.15em/\kern-.15em%
  \lower.25ex\hbox{\the\scriptfont0 00}}

\fi

\begin{document}
\thispagestyle{empty}



\markboth{{\sl \hfill  \hfill \protect\phantom{3}}}
        {{\protect\phantom{3}\sl \hfill  \hfill}}

\color{yellow} 
\hrule height 10mm depth 10mm width 170mm 
\color{black}

 \vskip -12mm   

\centerline{\bf \Large Deciphering the fluctuations of 
high frequency birth rates}
\vskip 5mm
\centerline{\bf \Large }
\vskip 10mm

\centerline{\large 
Claudiu Herteliu$ ^1 $, Peter Richmond$ ^2 $ and Bertrand M. Roehner$ ^3 $
}

\vskip 10mm
\large

%
{\bf Abstract}\qL
Here the term ``high frequency'' refers to daily, weekly or monthly
birth data.
The fluctuations of daily birth numbers show a succession
of spikes and dips which, at least at first sight, 
looks almost as random as white noise. However in recent times
several studies were published, including by the present authors,
which have given better insight into how birth is affected
by exogenous factors. One of them concerns the
way adverse conditions 
(e.g. famines, diseases, earthquakes, heat waves)
temporarily affect the conception capacity of populations,
thus producing  birth rate troughs 9 months after
mortality waves.  
In addition, religious interdicts
(e.g. during the Lent period) lead to reduced conceptions.
These as well as other effects 
raise the hope that we will soon be able to ``read''
and interpret birth rate patterns just as 
the Egyptologist Jean-Francois
Champollion managed to decipher many (though not all)
hieroglyphs.

\vskip 10mm
\centerline{\it \small Version of 20 February 2018}
\centerline{\it \small Provisional. Comments are welcome.}
\vskip 5mm

{\small Key-words: birth, fluctuations, holidays, religion, 
mass mortality, heat waves} 

\vskip 5mm

{\normalsize
1: Department of Statistics and Econometrics, Bucharest University
of Economic Studies, Bucharest, Romania. 
Email: claudiu.herteliu@gmail.com \qL
2: School of Physics, Trinity College Dublin, Ireland.
Email: peter\_richmond@ymail.com \qL
3: Institute for Theoretical and High Energy Physics (LPTHE),
University Pierre and Marie Curie, Sorbonne Universit\'e,
Centre de la Recherche Scientifique (CNRS).
Paris, France. \qL
Email: roehner@lpthe.jussieu.fr
}

\vfill\eject

\qI{Introduction}

That the
determinants of conceptions and births are still 
not well understood
is shown fairly clearly by the fact that 
measures taken to boost birth rates by
governments of countries experiencing
shrinking populations (e.g. Japan or Singapore)
proved largely ineffective. 
\qpar

The present paper is
not focused on such long-term changes but rather on
short- and medium-term changes. 
Nevertheless, some of the effects
studied here may also have long-term implications.
As an illustration one can mention the effect on births of
religious interdicts.
So numerous were such interdicts in
Middle Age western Europe
that only about 70 days were left
annually for permitted sexual relations. Had they been strictly
followed, such rules would have drastically restricted
fertility.
\qpar

Fig. 1 represents daily birth numbers in France in 1969.
This fast succession of upward and downward spikes resembles
a graphical representation of white noise except that instead of being
stationary the signal displays also some medium-term
ups and downs.
\qpar
%
\begin{figure}[htb]
\centerline{\psfig{width=12cm,figure=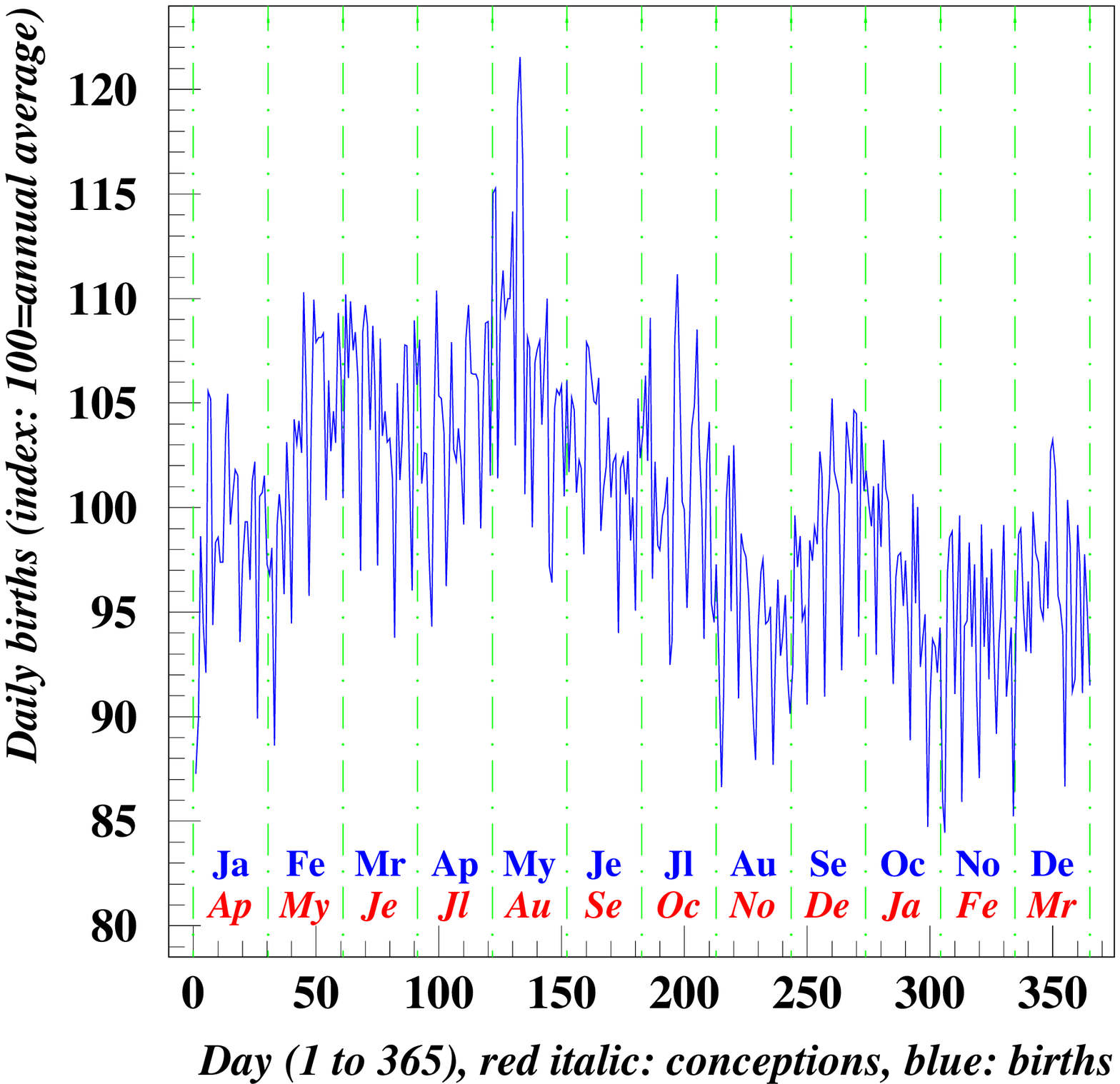}}
\qleg{Fig.\qhu 1\qhv Daily births in France in 1969.}
{The data were normalized by dividing them by the
annual average.}  
{Source: R\'epartition quotidienne des naissances
vivantes, France m\'etropolitaine. Institut National
de la Statistique et des \'Etudes \'Economiques (INSEE).
[Daily distribution of live births. Overseas territories
are excluded. National Institute for Statistics and
Economic Studies]. Available on Internet under the title:
``Les naissances en 2016. Tableaux de s\'eries longues''.}
\end{figure}
%

To make sense of such a jerky graph seems a real challenge.
Usually, the standard approach is to perform a moving window
average which will eliminate high frequency changes.
The remaining ups and downs will then be interpreted
in terms of specific human activities (e.g. vacation,
harvesting in rural societies and so on) 
which are assumed either to favor or reduce conception rates.
Most of the time such interpretations remain fairly qualitative
(see for instance Houdaille 1979, 1985)
for it is hardly possible to know in a quantitative
way the effect vacationing or harvesting has on the
sexual behavior of people. 
\qpar

The method that will be used here is different in the sense that we
will focus our attention on {\it sharply defined events} 
such as epidemics,
earthquakes, hot weather days, national or religious celebrations,
in an attempt to identify their
consequences in terms of birth rate changes nine months later.
In other words, we are not interested in ordinary life
activities (such
as vacationing or harvesting) but rather in exceptional events.
In this way we will be able to build a kind of dictionary
listing the connections between specific peaks or
troughs
and the events which triggered them 9 months earlier.
As implied by the parallel with Champollion made in the
abstract, this dictionary will not give the clue
for all accidents observed in birth series. Unexplained spikes
or troughs
may point the attention of researchers to
new, still unknown, phenomena.
\qpar

The paper proceeds through the following steps.
\qee{1} First, we identify daily spikes attributable to holidays
during which medically induced births are drastically reduced.
In this way, all national holidays which occur on fixed days
(such as 4 July in the US or 14 July in France) will become
apparent. This is a fairly easy step which only
requires an averaging procedure over a sufficiently large
number of years. 
\qee{2} Secondly, we focus on adverse conditions which create
troughs nine months later. This part relies on the results given
in a recent paper (Richmond et al. 2018a).
\qee{3} The two previous parts give interpretations of dips
and narrow troughs but not of wide troughs nor of peaks.
Here we show that broad troughs can result from
religious interdicts and that birth peaks may be brought about
by religious celebrations.

\qI{``Days off'' effect}

\qA{Identification of the cause of the downward spikes}

Fig. 2a shows the curve that one obtains by summing up
35 annual curves (restricted to non-leap years)
such as the one represented in Fig. 1.
Not surprisingly, the dispersion is reduced. Whereas
the curve of Fig. 1 moves within $ 100\pm 10 $
the curve of Fig. 2 moves within $ 100\pm 5 $.
\qpar
If successive years were independent their average 
would have a dispersion 6 times (i.e. $ \sqrt{35} $)
smaller than the initial series. The fact that the dispersion
is only divided by 2 shows that the successive series
are highly correlated; this point was already discussed
in Richmond et al. (2018b). 
%
\begin{figure}[htb]
\centerline{\psfig{width=12cm,figure=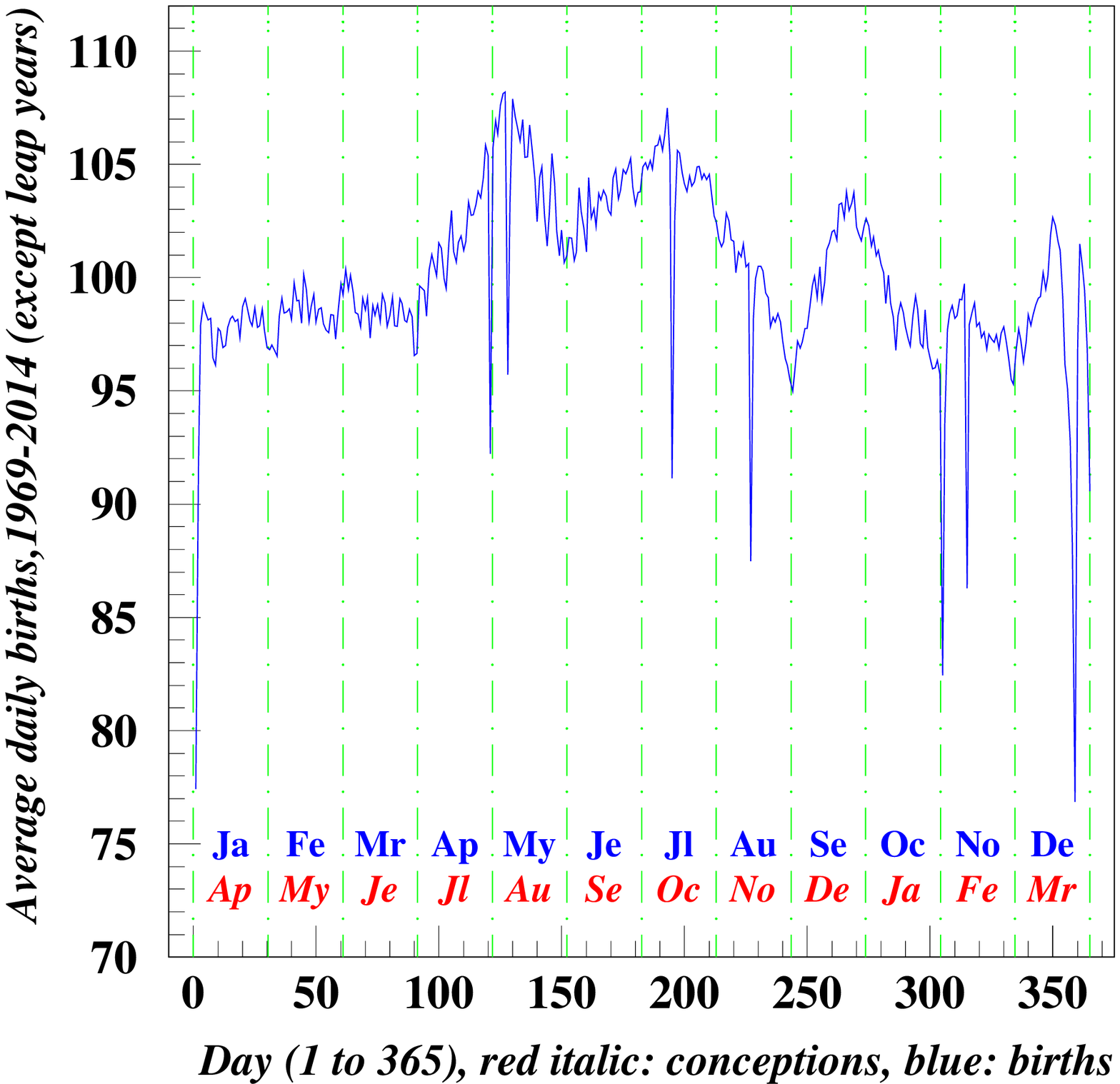}}
\qleg{Fig.\qhu 2a\qhv Average of daily births in France 
over 35 years, 1 January=1}
{The data cover 1969--2014 (leap years were excluded).
The downward spikes correspond to non-mobile public 
holidays in France: 1 May is Labor Day, 8 May is end of World
War I, 14 Jul is Bastille Day, 1 Nov is in memory of
all deceased, 11 Nov is end of World War I, 24-25 Dec
is Christmas.}  
{Source: Same as for Fig.1.}
\end{figure}
%

%
\begin{figure}[htb]
\centerline{\psfig{width=12cm,figure=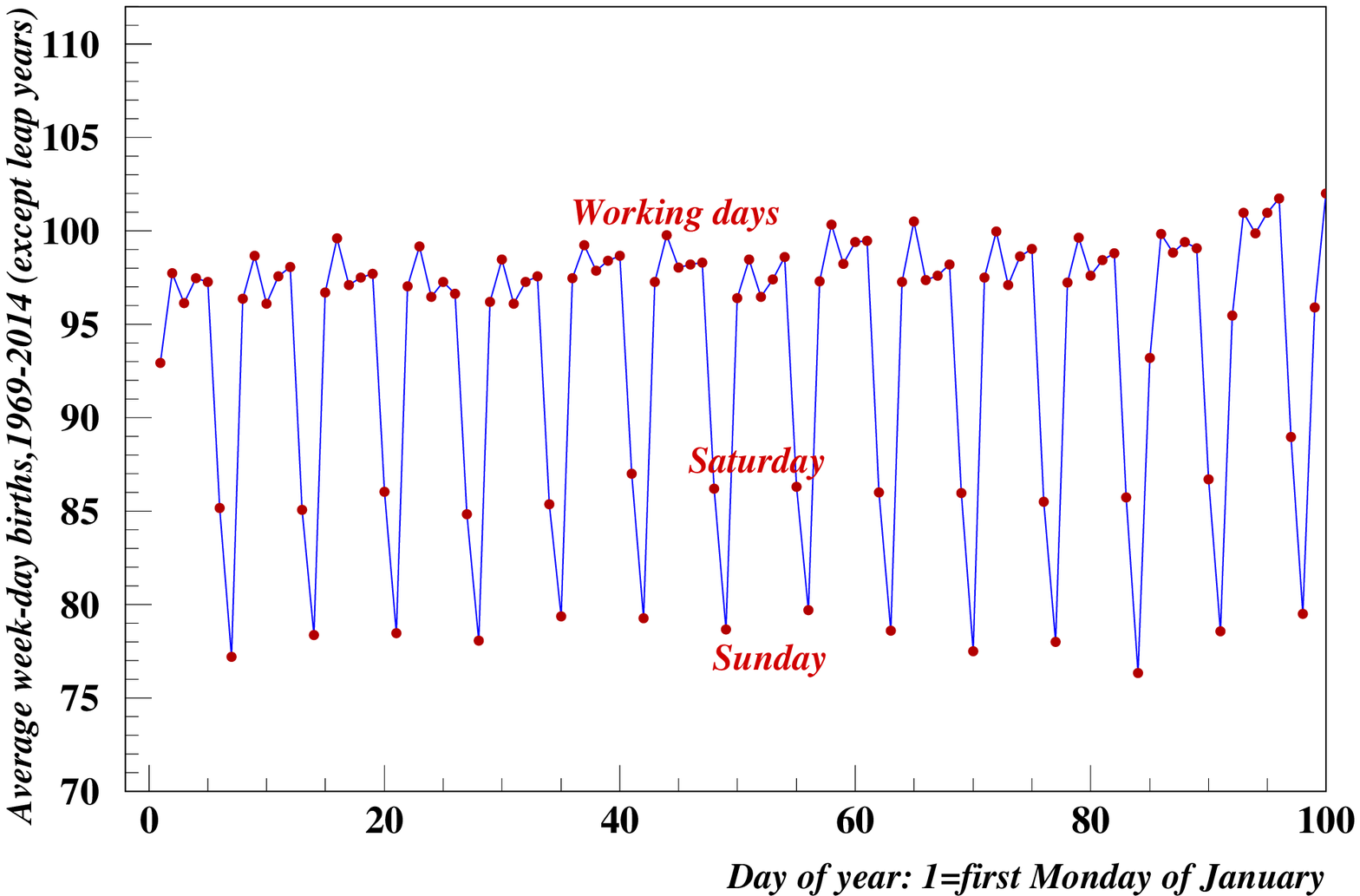}}
\qleg{Fig.\qhu 2b\qhv Average of daily births in France 
over 35 years starting with first Monday of January.}
{The data cover 1969--2014 (leap years were excluded).
The downward spikes correspond to Saturdays and Sundays.
It is for the sake of clarity that
the series was limited to 100 days (instead of 51 weeks).}  
{Source: Same as for Fig.1.}
\end{figure}
%

However, there are downward pointing spikes which reach
well beyond this interval. From their positions it is 
easy to recognize that they correspond to fixed public
holidays. The reduction in the number of births on non
work days relates to
medical induction of labor where drugs are used
to induce labor in advance of it beginning naturally.
Obviously such interventions will rather be planned on
working days. In 2014 some
23\% of American pregnant women experienced labor induction
up from a percentage of 10\% in 1990.
Cesarean delivery has also increased: from 15\% in 1996 to
some 22\% in 2011. Note the figures for induced labour and
Cesarian delivery should not be added together since the former
may lead to the latter.
\qfoot{The sources are: 
``Quick facts about labor induction'', August 2016;
Centers for Disease Control: Recent declines in induction of labor by
gestational age, June 2014.}%
.
The proportion of induced delivery is about the same in France
(Enqu\^ete nationale p\'erinatale [Perinatal national survey] 2010).
\qpar
For the period of 35 years following 1969
(excluding leap years) 
the average reduction in birth numbers for public holidays
is about 12\%.

\qA{Effect of Saturdays and Sundays}

On Fig. 2a one cannot clearly see the effect of mobile public holidays.
Examples are Easter or  ``Labor Day'' in the US which is
on the first Monday of September. For the same reason, the
effect of Saturdays and Sundays does not appear. 
However, the weekends appear in Fig.2b.
Between Fig 2a and 2b there is only a small difference
in design but it makes a drastic difference in shape.
In Fig.2a, because the annual series start at 1 January
which can be any day of the week
the weekends are positioned fairly
randomly with the result that in successive years
the summation mixes weekends and working days.
On the contrary  in Fig.2b the annual series
start at the first Monday of January with the 
result that the weekends
are at the same locations in successive years; thus,
the summation singles them out in a clear way.
By direct
examination of the data, one finds the following reductions
in birth numbers for 1969 and 2016.
%
\begin{table}[htb]

\small
\centerline{\bf Table 1: Effect of Saturdays and Sundays on birth
numbers with respect to working days.}

\vskip 5mm
\hrule
\vskip 0.7mm
\hrule
\vskip 0.5mm
$$ \matrix{
\qtb
\hbox{Year} \hfill &\hbox{Saturday} & \hbox{Sunday} \cr
\noalign{\hrule}
\qth 
1969 & -1.2\% & -6.8\% \cr
\qtb
2016 & -15\% & -19\% \cr
\noalign{\hrule}
} $$
\vskip 1.5mm
Notes: The data are for France. The reduction indicated in the table
refers to the difference between Saturday or Sunday and the
average of adjacent working days. \qL
Sources: Same as for Fig. 1. The topic of weekend birth
rates is also studied and discussed in Lerchl (2005),  
Lerchl et al.(2008) and Lerchl (2008).
\vskip 2mm
\hrule
\vskip 0.7mm
\hrule
\end{table}
%
The increase in the reduction  reflects the
increase in the proportion of induced and Cesarean deliveries.

\qA{Removal of the ``Days off'' effect}

In order to estimate the global effect that days off have
on birth numbers, we replaced these data by the average of
the births observed on
adjacent working days. The resulting shape (see Fig. 3) 
is much less jerky than in Fig. 1. 

%
\begin{figure}[htb]
\centerline{\psfig{width=12cm,figure=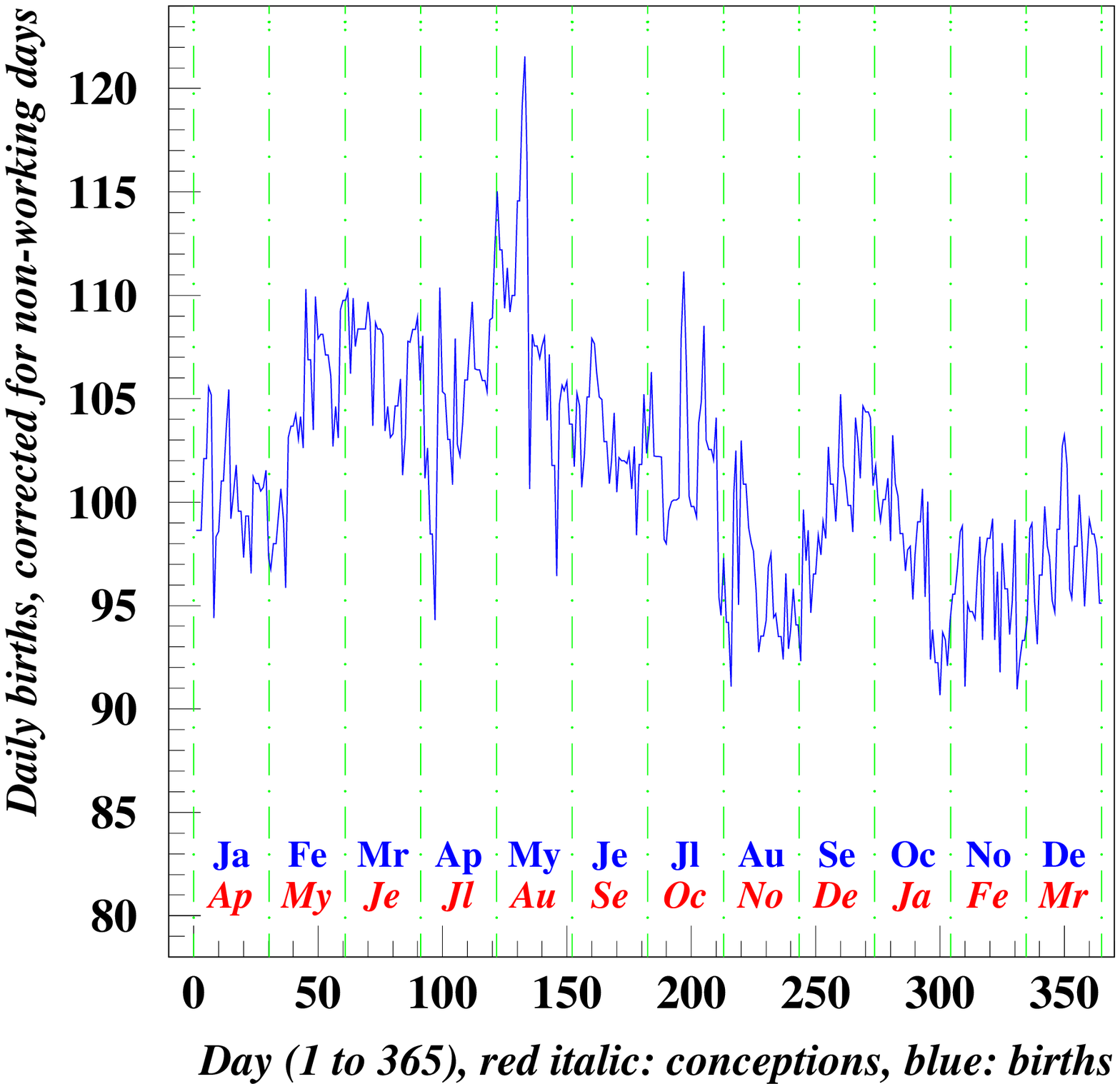}}
\qleg{Fig.\qhu 3\qhv Daily births in France in 1969
after anomalous non-working day data were smoothed out.}
{Data for days off were replaced by the average of the
births on the two adjacent working days. This resulted in a reduction
of the standard deviation from 6.12 in Fig. 1 to 5.29 here.}  
{Source: Same as for Fig.1.}
\end{figure}
%
The removal of the days off was performed in 3 steps which resulted
in the following reductions of the standard deviation:
$$ 
\hbox{\normalsize Initial: }\sigma_1=6.1,\ 
\hbox{\normalsize No public holidays: }\sigma_2=6.0,\
\hbox{\normalsize No Sundays: }\sigma_3=5.4,\
\hbox{\normalsize No Saturdays: }\sigma_4=5.3 $$

\qI{Effect of adverse living conditions on conceptions}

The removal of the ``Days off'' effect  has a dramatic 
impact on the shape of the curve of daily births, but one must
recognize that it is a fairly obvious effect. Less trivial
is the effect resulting from adverse living conditions 
which most often are revealed through a surge in mortality.
This effect is less obvious for two reasons.
\qbu It occurs at time of conception that is to say 9 months
before the births.
\qbu From conception to birth there are several successive steps
which involve social as well as biological phenomena.
It is a comparative analysis of several case studies which
revealed that the key-role is played by
social rather than biological factors (Richmond et al. 2018a,b).
\qpar

In this section, this effect will be presented through two
cases.
\qee{1} The impact that influenza outbreaks have on conceptions.
\qee{2} The impact that heat waves have on conceptions.

\qA{Implication of influenza outbreaks for conceptions}

In Richmond et al. (2018a) the authors used data for influenza outbreaks
of exceptional gravity such as in 1889 or 1918. Here we will
consider ``ordinary'' outbreaks of the kind that occur every
year either in December or (most often) in January. Naturally,
the effect is less spectacular but this analysis will show
that even small outbreaks have a sizable effect on conceptions.

%
\begin{figure}[htb]
\centerline{\psfig{width=16cm,figure=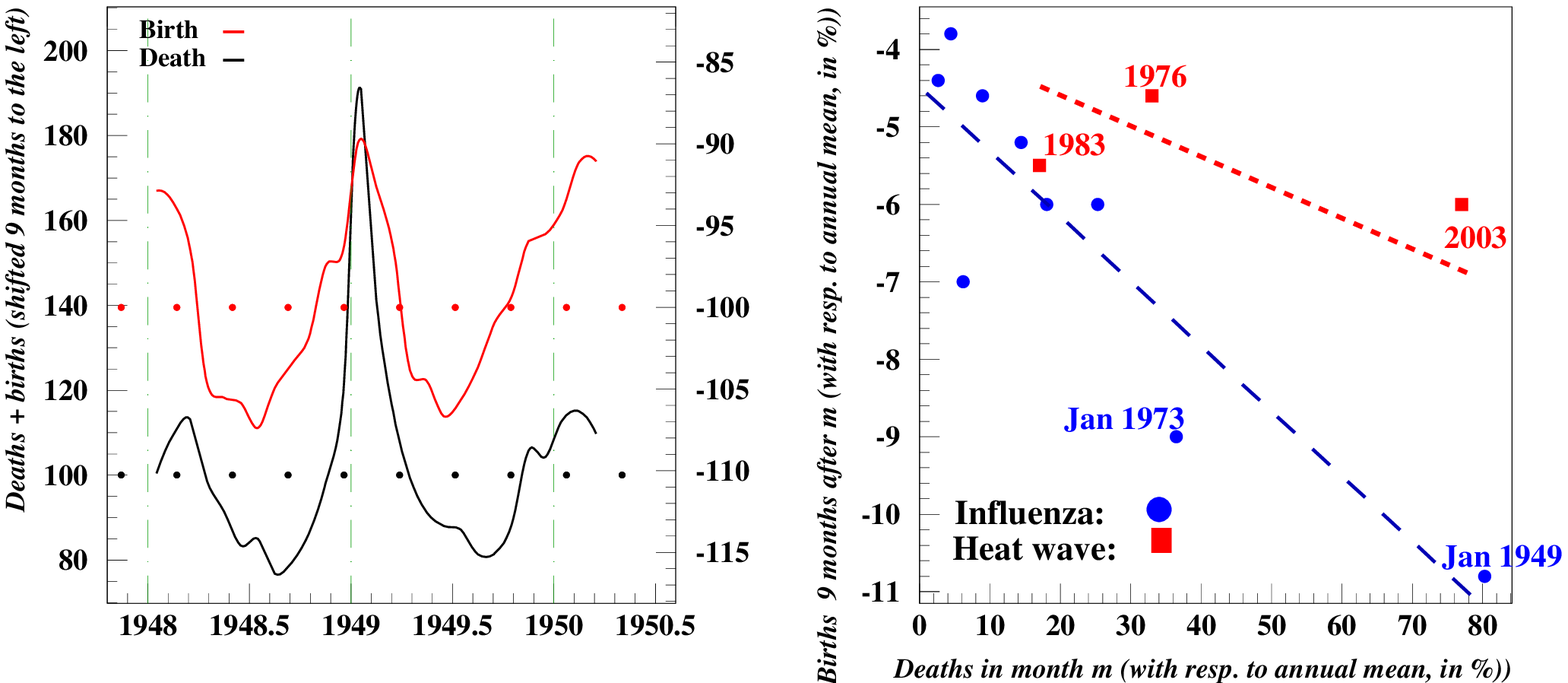}}
\qleg{Fig.\qhu 4a,b\qhv The effect of influenza outbreaks on
births in France between 1948 and 1974.}
{{\bf Left}: Relationship between monthly deaths and monthly births
nine months later. {\bf Right} Relationship between monthly
excess deaths in January (with respect to the annual average
and reduction in births in October of the
same year. The coefficient of correlation is $ -0.89 $ and
the equation of the regression line is: $ y=ax+b, a=-0.084\pm 0.03,
b=-4.5\pm 0.7 $. The three squares in red correspond to heat waves
in France (see the text).} 
{Sources: Births: Same website as for Fig. 1. Deaths: ``Nombre
moyen de d\'ec\`es par jour selon le mois depuis 1946 pour la France
m\'etropolitaine'', INSEE. [Monthly numbers of deaths divided by
the length of each month from 1946 to 2016]. Available on Internet
under the title: ``Les d\'ec\`es en 2016. Tableaux de s\'eries
longues''. } 
\end{figure}
%

In Richmond et al. (2018a) the focus was on 
special events which means that
we knew in advance the year and month we should consider.
Here, on the contrary, there are no special events; thus,
how do we know the years and months on which to focus?
The following answer relies on the fact that the influenza outbreaks
are not the same every year, despite of the fact that their timings
are fairly similar.
\qbu For the month, there is little hesitation for one knows
that it is January which is the peak month for influenza deaths.
It is true that in some years the peak is in December or in February
but even in such cases (which are uncommon, less than 20\%)
January also has a high death rate (Fig. 7a,b show 
death rates in the decades
1946--1955, 1996--2005 in France, see also the
CDC Wonder database for 1999--2016 in the US).
\qbu As to the year, it should be observed that there are in
fact differences from year to year. The strategy here is to
select the years which have the highest January death rates.
In this respect it is important to observe that there is a
high degree of correlation between January death rates either
for influenza or for all causes of deaths.\qL
For instance, in the United States in the 18 years from
1999 to 2016 the correlation is as high as 0.82%
\qfoot{For the percentage deviations with respect to 
annual averages the regression reads: $ y=ax+b, a=5.9\pm 2, b=\pm 4 $,
where $ x $ represents excess influenza deaths and $ y $ 
excess deaths for all causes. The data used here are from the
CDC-Wonder database.}%
.
This is a remarkable result for influenza and pneumonia 
(code J09-18 of the International Classification of Diseases)  
deaths in fact represent
only a small fraction (about 3\%) of the total deaths of January.
The fact that despite this small fraction influenza nevertheless
controls the January deaths 
is due to the fact that, apart from influenza changes,
the total January deaths are very stable; their coefficient of variation
is only about 5\%. Thanks to the high correlation, we can select the
appropriate years by using either the monthly influenza deaths 
or, if those data are not available, the monthly totals.  
\qpar

In Fig. 4b we selected the years in the interval 1949-1974
for which the January deaths were higher than the 
annual average. The plot
shows clearly that there is a distinct impact on births
for the highest death peaks which are the analogues of the
exceptional events studied in Richmond et al. (2018a).
For low January death rates, the impact of influenza deaths
becomes masked by the background noise. In other words,
in such cases the effect may still exist but it is no longer visible.
\qpar
In the rest of the paper, for the sake of brevity, 
the effect described here will be called the 
{\it death-nobirth} effect.

\qA{How deaths are affected by heat waves}

It is a fairly common sense notion that heat waves may affect
fragile persons particularly babies and elderly people.
Needless to say,
one would expect this effect to be more pronounced in
countries (like France or Switzerland considered below)
where air conditioning is fairly uncommon.
Although there have been many papers which document this effect
only few give age-specific results. Among the later, there are
two fairly recent papers which are
particularly commendable in this respect, namely Rey et al. (2007)
for France
and Vicedo-Cabrera et al. (2016) for Switzerland. 
\qpar
From our present perspective
it is particularly important to know what age groups are
mostly affected for, if only babies and elderly are affected, there
will be no reduction in births nine months later.
In Rey et al. (2007) there is a very interesting table which gives
all the information that we need.

%
\begin{table}[htb]

\small
\centerline{\bf Table 2: Age distribution of the victims 
of heat waves and influenza respectively.}

\vskip 5mm
\hrule
\vskip 0.7mm
\hrule
\vskip 0.5mm
$$ \matrix{
\qth
&\hbox{All ages} &\hbox{Age}  &\hbox{Age} & \hbox{Age} \cr
\qtb
&0-100&0-35 & 35-75 & 75-100\cr
\noalign{\hrule}
\qth
\hbox{Deaths during heat waves, av. 1975--2003 (1)} \hfill & 4,400 & 3.5\% 
& 24\% & 70\% \cr 
\hbox{Deaths in normal conditions, av. 1975--2010 (2)} \hfill & &8.0\% &
37\% & 54\% \cr 
\hbox{Excess mortality ratio (1)/(2)} \hfill & &0.33 & 0.65 & 1.30 \cr
\hbox{} \hfill &  &  &  \cr
\hbox{Deaths from influenza, av. 1999--2016 (1)} \hfill & 57,700&1.8\% & 
 23\% & 75\% \cr 
\hbox{Deaths in normal conditions, 1999-2016 (2)} \hfill & &4.5\% &
39\% & 56\% \cr 
\qtb
\hbox{Excess mortality ratio (1)/(2)} \hfill & &0.40 & 0.59 & 1.34 \cr
\noalign{\hrule}
} $$
\vskip 1.5mm
Notes: The heat wave data are for France and are an average of 6 heat waves,
namely: 1975, 1976, 1983, 1990, 2001, 2003. The influenza data are
for the United States and are an average for 1999--2016.
Taking into account that the US population is about 5 times
larger than the French, we see that for a same population
there are about 2.6 times more influenza deaths than heat wave deaths.
Moreover, contrary to influenza outbreaks that occur every year 
heat waves take place on average only every 5 years.
In terms of gender distribution, for heat waves
females represented 58\% of the victims, a figure which is not
surprising because females are over-represented in the age group
over 75 years.\qL
Source: Heat waves: Rey et al. (2007, Table 1). Influenza: CDC-Wonder
database.
\vskip 2mm
\hrule
\vskip 0.7mm
\hrule
\end{table}
%
%

Table 2 shows that, not surprisingly, 
young people under 35 are under-represented among heat wave
victims
while elderly people over 75 are over-represented. However,
the same table shows that for influenza the proportions are almost 
the same. 
Actually, we are not interested in the persons 
who die but rather in those who are affected and survive.
Thus, the real question is: ``Are influenza survivors more
affected than heat wave survivors?''
By studying the effect on births, we can answer this question
at least in its specific meaning of ``being affected in 
terms of conception capacity''. This is the point to which we come 
now.

\qA{How conceptions are affected by heat waves}

Whereas there have been many papers about the effect of heat 
waves on death rates, to our best knowledge there was only
one author who investigated its effects on birth rates nine
months later. In a pioneering study
Arnaud R\'egnier-Loilier analyzed
the effect on conceptions and births of three 
heat waves that occurred in France, namely: 1976, 1983 and
2003. Here the percentages of deaths with respect to 
annual deaths are of the order of a few percent that is to say
about 10 times smaller than for influenza deaths.
This makes the analysis more tricky. In order to get rid 
of the seasonal fluctuations R\'egnier-Loilier divided the
births of the year under consideration by the series of the
two adjacent years. This leads to the identification of
birth troughs  whose width is about 40 days.  
\qpar
Such shocks belong to the same death-nobirth family as those 
for famine, influenza or earthquakes considered 
above and in Richmond et al.
(2018a). That is why it makes sense to plot the results in the
graph of Fig. 4b. We can see that for the same percentage of deaths
one gets smaller conception troughs. This provides an answer to
the question raised above. We see that (for the same number of
fatalities) the heat wave survivors are less affected than
the influenza survivors, a rather natural conclusion intuitively.
\qpar

In conclusion to this subsection about
the effect of heat waves, we wish to point out some
remaining uncertainties and interrogations.
\qbu The identification of the troughs is somewhat doubtful
because, except in 2004, 
they are not visible on the birth time series: 
they do not appear on the monthly data and, 
due to their large fluctuations, even less on the daily data.
They become visible only after two operations have been performed.
First, each annual birth series $ b $ is ``renormalized''
through division by the average of the two adjacent series,
e.g. $ b=(1977) $ is replaced by 
$ b'=1977/[(1976+1978)/2] $.
One would expect such a division to reduce the variability
(that should be so if the denominator is sufficiently
similar to the numerator)
but that is not what is observed: actually in the three cases
of 1977,
1984, 2004, the coefficient of variation of the initial series
is (on average) smaller than the CV of the renormalized series
(the ratio is 0.97). \qL
The second operation is a broad moving
average with a window width of 21 days. These operations lead to
series which have more ups and downs than the
initial series would have after a similar moving average.\qL
In short, the effect may well exist but due to its small
magnitude it is at the limit of being detectable.
\qbu The second interrogation concerns
the specific mechanism which brings about the birth troughs.
Is it a biological or a behavioral effect? In other words,
does the higher temperature affect the conception mechanisms
or does it discourage intercourse? The high fertility
that one sees in many African countries calls in doubt a
biological incidence. The fact that similar birth reductions
are observed in the wake of earthquakes also weights in
favor of the behavioral explanation.

\qA{Removal of the death-nobirth effect for daily births}

Now we wish to see how the fluctuations of births 
shown in Fig.3 are
reduced when the death-nobirth effect is removed, that is to say when
the dips due to excess-deaths 9 months earlier are compensated for.
This compensation process is done by using the linear regression
equation given in the the caption of Fig. 4b.
The standard deviation is reduced from the previous
value of $ \sigma_4= 5.3 $ to $ \sigma_5=4.7 $ (a reduction of 11\%).
This is a fairly small reduction and Fig. 5 explains why it is not
larger. 

%
\begin{figure}[htb]
\centerline{\psfig{width=12cm,figure=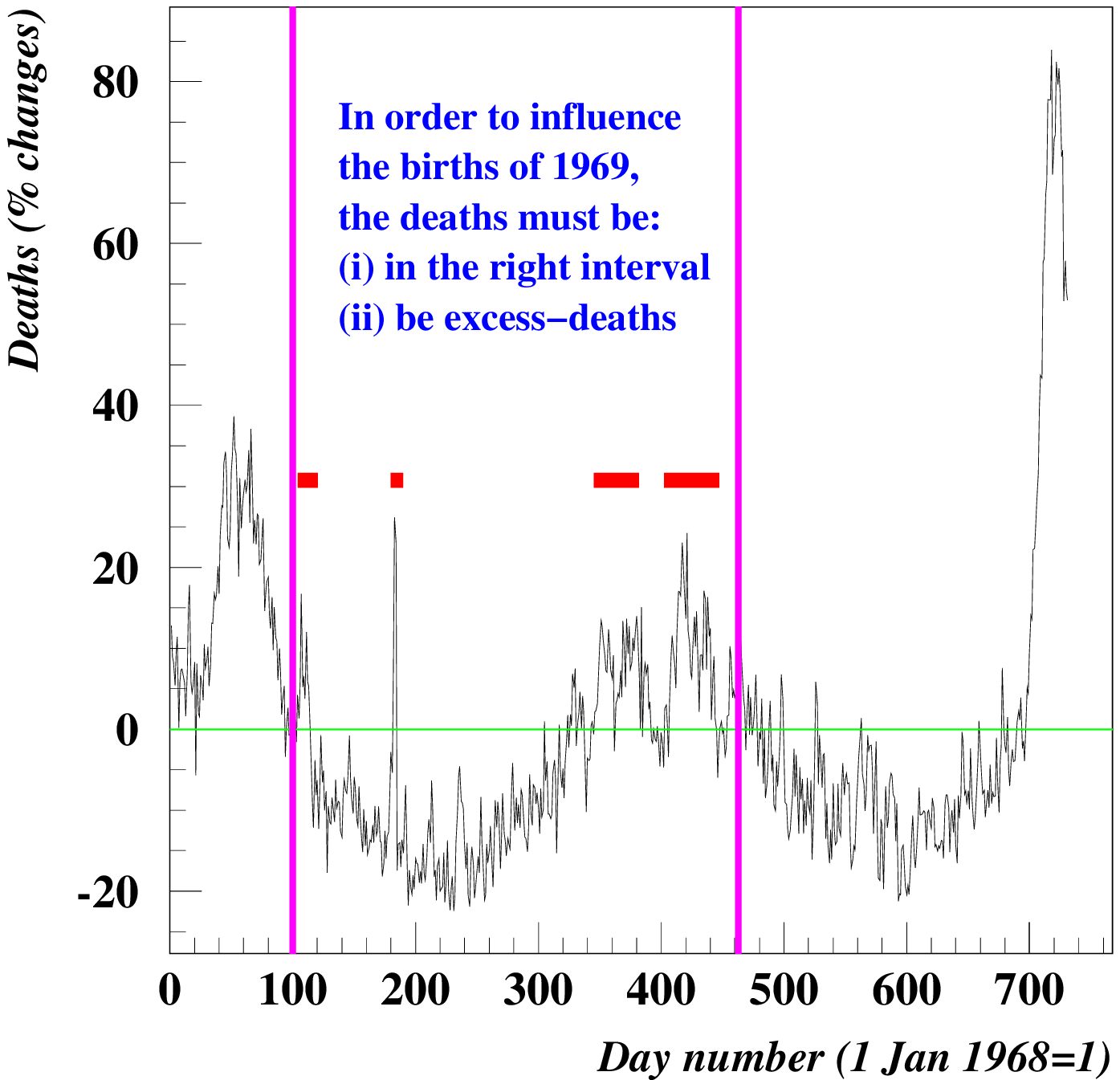}}
\qleg{Fig.\qhu 5\qhv Removal of the death-nobirth effect,
for daily births, France 1968--1969.}
{This figure explains why in 1969 the removal of the death-nobirth
effect has only a small incidence.}   
{Sources: Same website as for Fig. 4.} 
\end{figure}

In order to affect the births of 1969, the deaths must
occur in the time interval shown in magenta in Fig.5.%
\qfoot{More precisely we have taken a pregnancy duration of 267 days
as explained in Richmond et al. (2018a).}%
;
within the two years 1968--1969 this defines a specific time interval.
It turns out that in this interval deaths are mostly
below average; there are only small 
sub-intervals (marked in red) which have excess-deaths. Therefore
the curve of the births during 1969 is only slightly changed.
The fact that despite this small impact the standard deviation of
births is nevertheless reduced by 11\% shows that the correction
works indeed in the expected way that is to say by pushing up
the troughs; the modified birth curve is omitted because
visually the changes are rather subdued.

\qA{Removal of the death-nobirth effect for monthly births}

Graphically
the effect of the removal of the death-nobirth effect will become
clearer at the level of monthly births. This is shown in Fig. 6.
 
%
\begin{figure}[htb]
\centerline{\psfig{width=16cm,figure=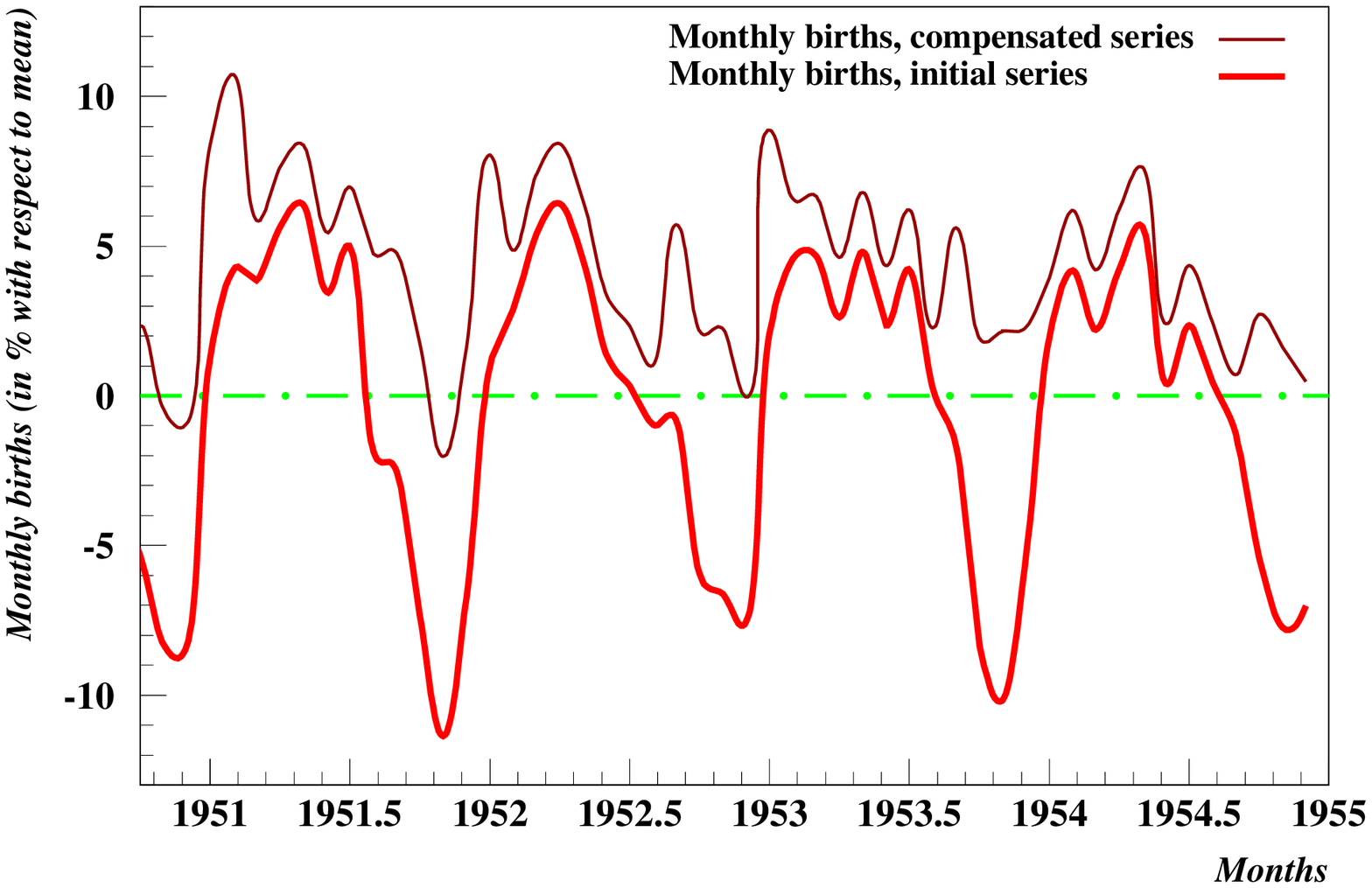}}
\qleg{Fig.\qhu 6\qhv Removal of the death-nobirth effect,
France 1950-1954.}
{For the compensation we used the regression coefficients $ a,b $
obtained in Fig. 4b; then the compensation was done as follows:
$ \Delta b=a\Delta d +b $, 
where $ \Delta d, \Delta b $ are the excess deaths and excess births
respectively; then $ b_c=b_i+\Delta b $ where $ b_i,b_c $ are the
initial and compensated birth series respectively.
Note that $ \Delta b $ is negative and that this compensation was
done only when there were excess deaths, that is to say when the 
death series was below its average. In the graph, the 
compensated series was translated somewhat upward
to avoid a superposition with
the initial series in all the intervals in which there are
no excess-deaths and therefore no compensation.}   
{Sources: Same website as for Fig. 4.} 
\end{figure}
%
It can be seen that all deep troughs occurring around September
are replaced by fairly shallow troughs. 
This corresponds to a reduction of the standard deviation from
5.11 for the initial series to 2.90 for the compensated series.
\qpar

The success of this procedure raises the question
of whether nevertheless 
there remain some deep troughs outside of the
time interval shown in Fig. 5. This point is discussed 
below; it will be seen that in recent decades a new
phenomenon appeared which brought about unexpected birth troughs

\qA{Beyond the death-nobirth effect}

Before considering a decade which shows birth troughs that
do not match expectations, let us show one 
in which the death-nobirth effect accounts for the
observed birth troughs (Fig. 7a).

%
\begin{figure}[htb]
\centerline{\psfig{width=13cm,figure=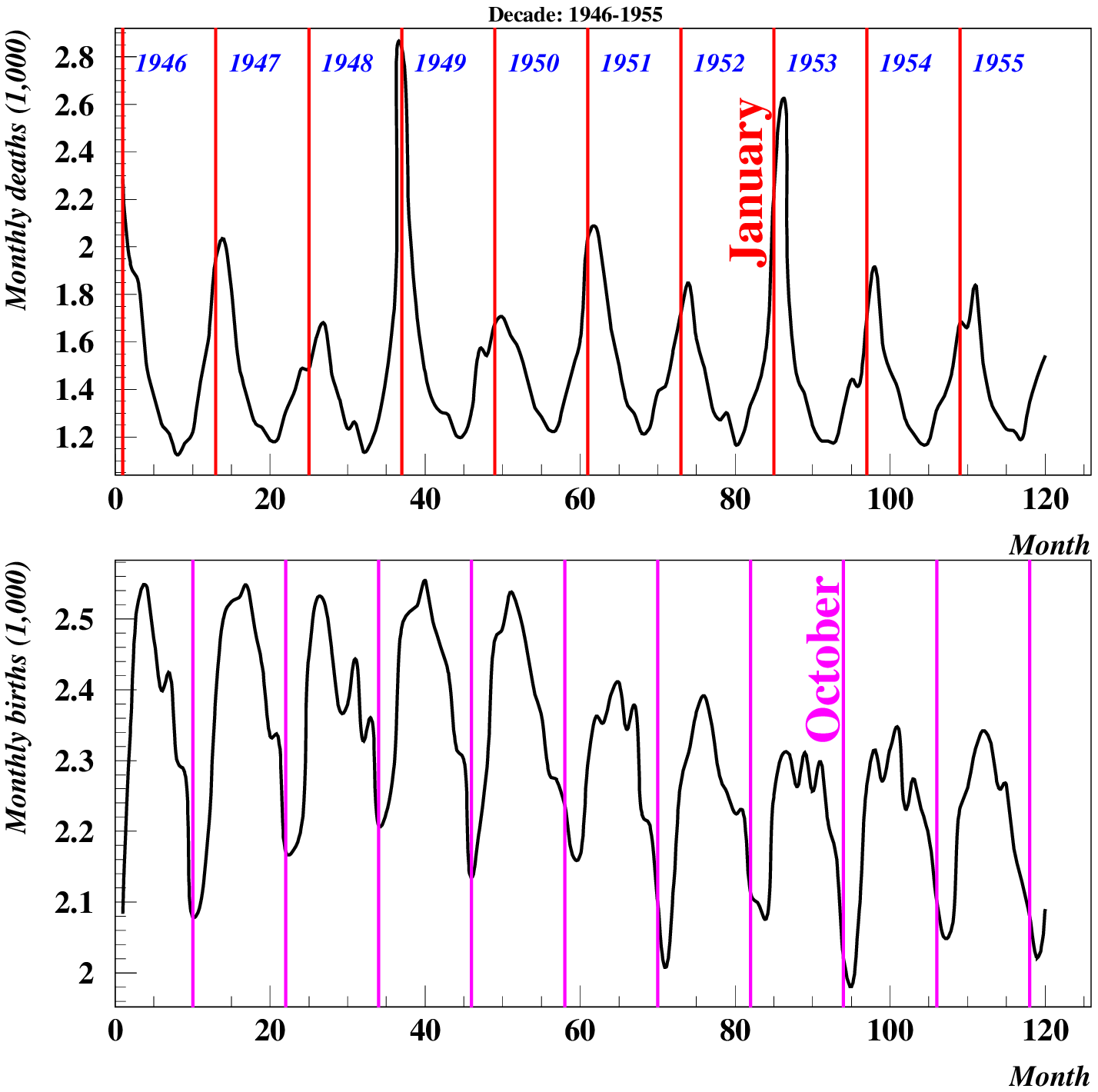}}
\qleg{Fig.\qhu 7a\qhv Illustration of the death-nobirth effect.}
{The data are for France and correspond to monthly deaths and births
divided by the number of days of each month.
All red lines of the death graph
indicate January and all magenta lines of the birth graph
indicate October (i.e. January+9months). It can be seen that
the birth troughs follow the death spikes with the expected 
time lag of 9 months. However, as shown in the next graph,
in the decades after 1990
the birth troughs moved progressively to the left
of October.}
{Source: Same website as for Fig. 4.} 
\end{figure}
%
%
\begin{figure}[htb]
\centerline{\psfig{width=13cm,figure=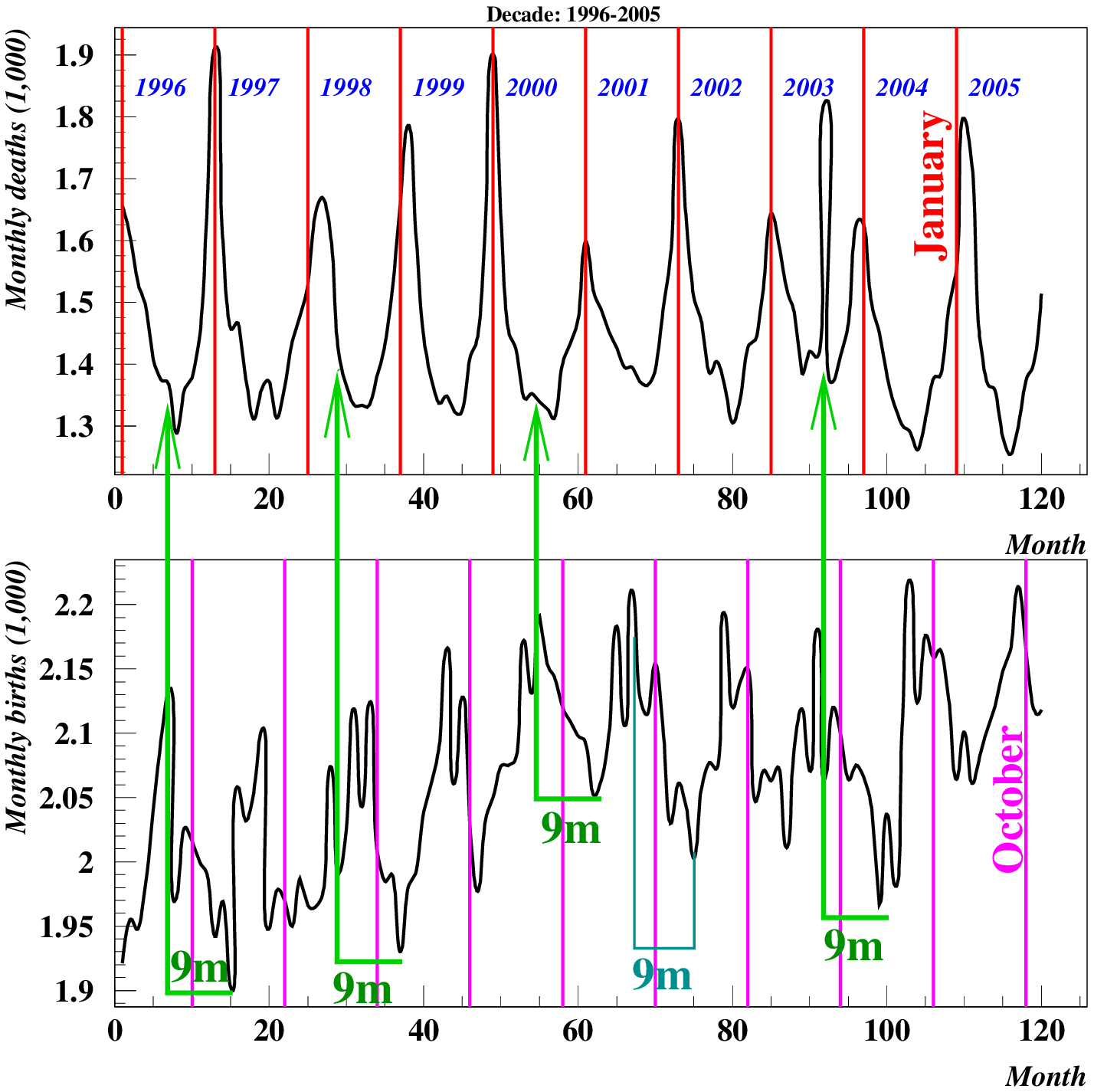}}
\qleg{Fig.\qhu 7b\qhv Apparition of a new effect which produces 
birth troughs not related to death spikes.}
{Here, in most of the cases the time lags between death spikes
and birth troughs are longer than 9 months. 
However the death spike associated with the heat wave 
of August 2003 is followed by a birth trough 9 months
later (i.e. in May 2004).
In addition, we observe a time lag close to 9 months
between birth peaks and birth troughs (see the 
U shaped line in cyan) for which an explanation is suggested
in the text.}
{Source: Same website as for Fig. 4.} 
\end{figure}
%
In Fig.7b we have birth troughs which follow a different pattern.
Actually it is a fairly loose pattern: in 1997 the birth
trough occurs in March, in 1998 there is no clearly defined
trough, in 1999, 2000, 2001  the troughs occur in January,
in 2002,2003 it is in February, in 2004 there is a double dip
in February and May one of which may be due to the death spike
of August 2003. In short, instead of the expected troughs
we see others that were unexpected.\qL
What explanations can one give?
\qbu More than the apparition of a new effect it is
the apparent disappearance of the death-nobirth effect
which is surprising.
In fact, it does not really disappear. 
The October dips are still present but
in an hidden way due to the presence of lower dips.
\qbu What can be the origin of the new troughs?\qL
In comparison with Fig.7a the births dips appear more
disorderly, often for instance displaying a double dip.
Similarly the birth peaks appear also more disorderly
with double dips often being preceded by double spikes.
This observation leads us to a possible explanation
which is based on the fact that for a couple
conception at time $ t $ excludes conception at a later
time at least until after birth. If all couples in
age of having a baby would conceive one in October,
then during the 9 subsequent months there would be no
conceptions. As a result, one would see a huge birth peak
in October + 9 months=July followed by a 9-month interval 
without any birth. In the real world where couples are
not highly synchronized one may expect a conception peak
to be followed by a conception trough 9 months later.
This is indeed what we observe (see the U shaped
line in cyan color). For the sake of brevity this
effect will be referred to as the {\it single conception} effect.\qL
Actually, it is quite possible that this effect contributed
in some way to the death-nobirth effect for in Fig.7a
one sees that the time interval between birth peaks and birth
dips is comprised between 6 and 9 months,
that is to say not too different from 9 months.\qL
A last question has to be answered, namely why did the
single conception effect suddenly overcome the death-nobirth
effect? A possible answer may be found in the width of the
birth peaks. Comparison of Fig.7a and 7b shows that they become
much more narrow which implies more synchronized conceptions,
therefore followed by deeper conception dips.

\qI{Influence of the religious environment on conceptions}  

There are religious precepts that rule sexual intercourse
during special periods. For instance, during the month of Ramadan,
sexual activity is not allowed during the day but is permitted
during the night. The orthodox religion advises against
sexual activity during Lent which in Orthodox countries
is a 7-week period (49 days) preceding
Easter. In present day Catholic and protestant religions
there are no specific precepts on this point. However,
although the scriptures do not say anything in this respect
in  past centuries the Catholic Church prohibited sexual activity
not only during Lent (40 days before Easter)
but also in many other periods,
e.g. 40 days before Christmas, 40 days after giving
birth, or on all the nights from Saturday to Sunday.\qL
Conversely, 
it can be mentioned that on some specific days,
sexual intercourse is encouraged by religious rules.
That is the case of Saturday (or more precisely Shabbat
which starts already on Friday night) in the Jewish religion.
\qpar

In connection with the present paper, we will address two questions.
\qee{1} Is there empirical evidence showing that some religious
precepts affected the pattern of births?
\qee{2}  Is it possible to document the impact of religious rules
in the course of time and across countries?

\qA{Empirical evidence: the case of the Fire Horse year in Japan}

Probably the easiest way to show the impact of beliefs on births
can be found in Japan. The reason which makes it easy is the
fact that one does not need to use monthly or even daily data.
Annual birth data are sufficient.
The population pyramid of Japan based on the census of 2000 presents a
mysterious discontinuity for people aged 34 that is to say who are born in
1966.
The number of people (both males and females) born in this year is much
smaller than in 1965 or in 1967. The difference is of the order of 30\%.
Further investigation reveals that 1966 was a Hinoeuma year, which means a
Fire Horse year in the Chinese calendar%
\qfoot{Because the Chinese New Year occurs in late January, the Fire Horse
Year does not exactly coincide with 1966; in fact, it started on 21 January 1966
and ended on 8 February 1967.}%
.
Girls born in that year grow up to be known as ``Fire Horse women'' and are
reputed to be headstrong and to bring bad luck to their families and to their
husbands. In 1966, as a baby's sex could not be reliably identified before
birth, there was a sharp fall in birth rate.
This fall was due partly to a fall in conceptions
and partly to an increase in abortions.  
According to the
Chinese calendar, Fire Horse years occur every 60 years; thus, the three 
previous one were in 1906, 1846 and 1786. It is a natural question to see
if the effects in those years were similar to the one in 1966.
For 1786 there are no data but for the three subsequent years
the answers are given in Table 3.
   
%
\begin{table}[htb]

\small
\centerline{\bf Table 3: Birth reductions in the last three Fire Horse years}

\vskip 5mm
\hrule
\vskip 0.7mm
\hrule
\vskip 0.5mm
$$ \matrix{
\qtb
\hbox{}  & 1846 & 1906 & 1966 \cr
\noalign{\hrule}
\qth \qtb
\hbox{Fall in annual births (\%)\hfill } & -11\% & -11\% & -24\% \cr 
\noalign{\hrule}
} $$
\vskip 1.5mm
Notes: The percentages refer to the differences between the year under
consideration and the mean of the 9 other years in the same decade.
The high reduction in 1966 is due to two factors.
(i) Contrary to what happened in 1846 and 1906, in 1966  the male-female
sex ratio at birth increased only by 1.3\% instead of 20\% in 1846
and 4.3\% in 1906;
this means almost no infanticide in 1966. 
(ii) In 1946 there was a sharp
fall in births which generated a small cohort; then, in 1966
this cohort reached the age of 20 at which point it started
to have children.                                            \qL
Source: Roehner (2007, p. 42).
\vskip 2mm
\hrule
\vskip 0.7mm
\hrule
\end{table}

\qA{Empirical evidence: the case of Lent in Romania}

Why should one focus on the case of Lent in Romania? There are
two good reasons for that.
\qbu As mentioned above, in contrast with the Catholic or
Protestant religions, the Orthodox religion has clearly
defined rules regarding sexual relations during Lent.
In the census of 2011, 86\% of the Romanian population
identified as part of the Eastern Orthodox Church.
\qbu Because the time of Lent changes from year to year
it is a much better ``marker'' than a fixed religious 
day. The reason is obvious. A single observation
may display a reduction in births but it will not tell
us the reason of this reduction.
On the contrary, if on successive years the birth
reductions follow closely the Lent time-intervals,
then there can be little doubt that Lent is the causal factor.
\qpar

The effect of religious rules on conception during Lent
was investigated in
Herteliu et al. (2015). For such a study one of the main difficulties
is the fact that strictly speaking monthly birth data are not
detailed enough to follow the shifts of Lent. One needs weekly
or daily data. Fortunately, the Romanian Bureau of the
Census had the clever idea to ask their birthday dates to
all persons surveyed in the censuses of 1992 and 2002.
\qpar

Naturally, the oldest persons may not remember their birthday
with absolute accuracy. In order to limit this uncertainty
the data set was limited to the persons born after 1905;
thus, the persons included in the data set were less than
87 year old in 1992. Because the size of a cohort decreases
rapidly after 65 the dataset would include only small
samples of persons born between 1905 and 1927. 
\qpar

For the whole period from
1930 to 2000 the reduction of conceptions during Lent
was found to be on average equal to: $ 14.1\% \pm 1\% $. \qL
In contrast there was almost no reduction during
the 40-day Nativity Fast (also called ``Advent'') that 
precedes Christmas.



\qA{Comparative perspective across countries}

On account of what we said above, one would expect
a smaller reduction effect in Catholic or Protestant countries
than in Orthodox countries
How can one test this conjecture?
\qpar

As daily or weekly birth data will not be available in the early
time periods that we wish to consider we made the following
simplified argument which will allow us to use monthly data.
\qee{1} Firstly, the comparison will be made on a time interval
before World War II, for instance 1931-1935 because
this will allow us to use the monthly data given in Bunle (1954).
\qee{2} Secondly, we observe that in western countries
Easter is on average on 8 April. Indeed,
averages performed over several 30-year intervals always
lead to the same date. Note that this is for
Catholic and Protestant countries.
For Orthodox Easter
the average date is somewhat later, namely on 21 April.
As in western countries
Lent lasts 40 days, on average western Lent will
last from 1 March to 8 April which means that if
we identify Lent with March we will miss only $ 8/40=20\% $.\qL
For Orthodox Easter the situation is less favorable.
As Lent lasts 49 days it will be
from 1 March to 21 April 
and will therefore be divided more equally (4/7 against 3/7) 
between March and April. By identifying Lent with March
we will miss $ 21/49=43\% $.
\qpar
There is an additional factor which makes the approximation
more acceptable.
If the religious injunction is followed fairly
closely one would expect a rebound of conceptions after Easter.
Such a rebound can indeed be observed; for instance in 
Bulgaria conceptions surged from 672 in March to 983 (+46\%) in
April. In western countries a surge can also be observed
but which is much smaller.
For instance in France from March to April conceptions
rise from 939 to 1042 (+11\%).
\qpar

In other words, in Orthodox countries
any reduction between 1 and 21 April might
be reduced
by a surge of conceptions in the rest of the month.
Thus, even in Orthodox countries,
March appears to be a more appropriate estimate of Lent
than April.
\qpar

If one accepts the previous argument the conception reduction
due to Lent can be represented by the ratio%
\qfoot{Because of the ``surge effect'' the conceptions of
April will be higher than they would be without such an effect.
As a result, $ R $ will be lower than it should be.}%
:\qL
\centerline{ \normalsize $ R= $
conceptions of March divided by one half of the
total conceptions of February and April}\qL
By adding 9 months this ratio will get translated into
an estimate for births. 

%
\begin{table}[htb]

\small
\centerline{\bf Table 4: Conception reduction during Lent,
1931--1935}

\vskip 5mm
\hrule
\vskip 0.7mm
\hrule
\vskip 0.5mm
$$ \matrix{
\qtb
\hbox{Conceptions in:} \hfill &\hbox{Feb. }(f)&\hbox{Mar. } (m)&
\hbox{Apr. }(a)& R=m/[(f+a)/2]\cr
\noalign{\hrule}
\qth
\hbox{Bulgaria}\hfill  & 891&688 &983 & 0.73 \cr
\hbox{Romania}\hfill  & 911&686 & 919&0.75  \cr
\hbox{}\hfill  & & & &  \cr
\hbox{France}\hfill  &985 &930 &1042 &0.95  \cr
\hbox{Spain}\hfill  & 911&947 &1096 & 0.94 \cr
\hbox{}\hfill  & & & &  \cr
\hbox{Norway}\hfill  &857 &899 &992 & 0.97 \cr
\qtb
\hbox{Sweden}\hfill  & 904& 941&979 &1.00  \cr
\noalign{\hrule}
} $$
\vskip 1.5mm
Notes: $ R $ equal to 1 would mean that March which represents
Lent has same conceptions as the two adjacent months
that is to say no reduction due to Lent.
The monthly numbers given in the table are an index
which sums up to 1,000 for one year. Bulgaria and Romania are
Orthodox countries, France and Spain are Catholic countries,
Norway and Sweden are Protestant countries.
\qL
Source: Bunle (1954, p. 92-93).
\vskip 2mm
\hrule
\vskip 0.7mm
\hrule
\end{table}
%

\qA{Comparative perspective across time}

One would expect a decrease of the reduction effect in the 
course of time. How can one test this conjecture?.

%
\begin{figure}[htb]
\centerline{\psfig{width=16cm,figure=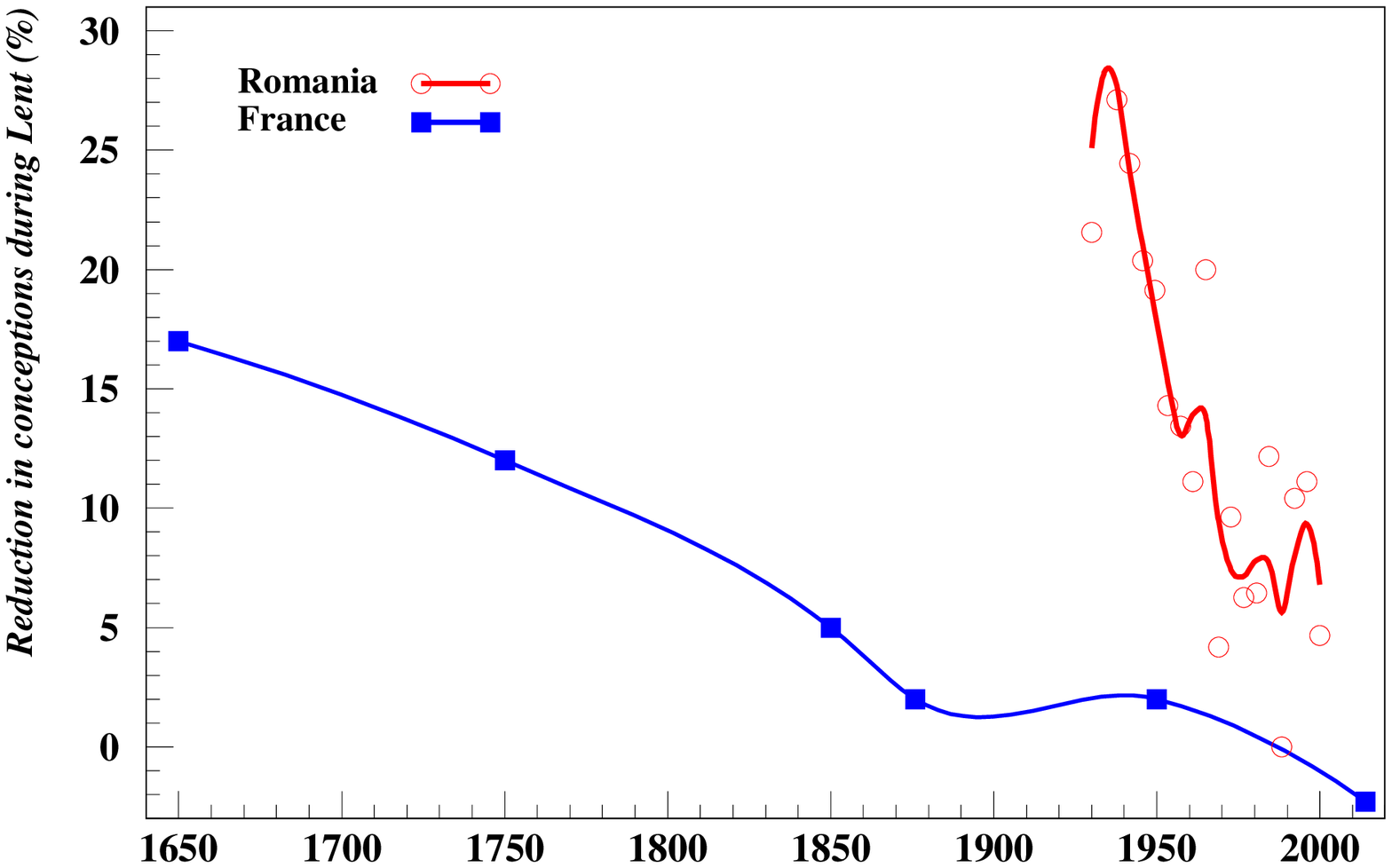}}
\qleg{Fig.\qhu 8\qhv Reduction of conceptions during
Lent in France and Romania.}
{The results for France are based on monthly birth data
whereas those for Romania are based on daily data which is why
they have a larger dispersion.
Actually, prior to the 19th century, the records 
give the dates of baptisms. In order to accept them 
as proxies of birth dates one must assume that 
baptism followed birth fairly closely. Overall the fall
of the Romanian curve is 6 times steeper than the fall
of the French curve.}
{Sources: Herteliu et al. (2015), Houdaile (1979, 1985),  
R\'egnier-Loilier 2010b.} 
\end{figure}
%

The data given in Herteliu et al. (2015) allow us 
to follow the reduction
from about 1930 to 2000. Needless to say, one would not
be surprised to see a fairly sharp fall of the impact
of religion during the 40 years of Communist regime.
However, in order to better assess the rapidity of the
fall we need to compare it with a non-Communist
country. If we use again the approximation which allows
us on average to identify Lent with March,
we can use French monthly birth data which cover
a broad time interval ranging from 1650 to 2000.
The comparison presented in Fig. 8 shows that 
in Romania the fall is about 6 times faster than 
it was in France.

\qI{Conclusion}

In this paper we have given interpretations of several
kinds of birth rate troughs. However, the origin of
birth rate peaks remains an open question%
\qfoot{There is a birth rate peak in late September (see Fig. 1,2,3 )
which  corresponds to a conception peak in December.
It may be due to the Christmas celebration.
However, how can one prove that it is really due to
Christmas rather than to some other factor?}%
.
\qpar

The main hallmark of the rules that we have studied is that,
to varying degrees,
they are valid in all countries for which we were able to
find reliable data. For instance, although in present-day
France the influence of religious beliefs is so small
that it can hardly be measured, the analysis of historical
data shows that in the past this effect did indeed exist.
\qpar

Obviously, this provides an objective way to test 
and assess the beliefs of people. We plan to extend this
investigation further as soon as new data become available.
For instance, when monthly birth data become available
for China it would be interesting to test if there are
conception surges in the ``Golden week'' (1-7 October)
or during the ``Spring festival'', that is to say the
week centered on the Chinese New Year.

\appendix

\qI{Appendix A: A historical view of previous studies}

Monthly birth data series became available in European countries
in the early 19th century and in the United States about
one century later (due to the fact that it took some time
to set up registration procedures in all states).
Explaining the monthly distribution
of births is a question which attracted the attention 
of researchers very early. 
One of the first studies was published in 1831
by Louis Ren\'e Villerm\'e (1782--1863), a French
physician and a pioneer of social epidemiology. In several respects
his article is typical of those published at that time and it
may be of interest to understand in what aspects it differs 
from present-day publications.
\qee{1} It is a long article of one hundred pages.
\qee{2} It is a {\it comparative study} which uses data from
many countries: France of course, but also Great
Britain, Germany (or rather
Bavaria, Prussia, Wurtemberg for at that time Germany was
not unified), Italy, the Netherlands, Russia,
Sweden. It uses also data from various time periods going
back for some places (e.g. Paris) to the 17th century.
This kind of comparative methodology remained in use during
the whole 19th century. Recall that
Durkheim's renowned study of suicide (1897) used exactly the
same methodology which, needless to say, was borrowed from
physics where experimenters also perform as many observations
as possible in order to understand any new phenomenon..
\qee{3} Interestingly, 30 pages of the paper are 
statistical tables which contain all the data used by the author.
Several of these data sets are still useful nowadays, as
for instance the monthly birth numbers for
Belgium and the Netherlands from 1815 to 1826.
The thanks expressed in the notes of the tables reveal that the
author had personal contacts with several foreign scholars
from whom he received many of his data.
\qee{4} In line with other scholars of the 19th century (e.g.
Alfred Espinas, 1878) Villerm\'e does not hesitate
to includes in his study
observations about animal species. This is of course in accordance
with the perspective of comparative analysis.
He observes for instance
that for foxes, hares and wolves conception takes place
between December and February which is fairly different
from humans for whom the conception peak is in spring.
\qee{5} Basically the paper states most of the general
rules we currently know about. In other
words, progress seems to have been fairly limited.
\qpar

Let us now mention some more recent studies,
e.g. MacFarlane (1974), Seiver (1985), Lam et al. (1994).
Huber et al. (2011).
They are of two types. The first three are fairly broad studies  
whereas the last one focuses on a very specific point.
For all of them one could say that they provide evidence
but no real understanding; actually
not even an ``observational understanding''%
\qfoot{By this expression we mean a rule 
valid in all countries where data are available
which gives it real predictive power.}%
This is fairly obvious for the specialized
paper because in this case the question it addresses is too specific
to be understood from any basic principles that may exist. 
The same is true for the broad papers yet for a different reason. 
Although they offer evidence, even
in some cases such as MacFarlane (1974) comparative evidence
for several countries,
they refrain from raising any questions. For instance, 
Lam et al. (1994) show that between the
monthlu birth patterns of 
Britain and the US state of Georgia there is a 6 month time
lag. This difference comes of course as a surprise for two
regions that are fairly similar in many respects. 
However the authors do not try to explain it.
Even if the reason cannot be identified, would it not be useful to 
point out that, at least for now, there is no satisfactory
explanation? This might stimulate more sharply targeted research.

\vskip 4mm

{\bf References}

\qparr
Bunle (H.) 1954: Le mouvement naturel de la population dans le
monde de 1906 \`a 1936. [Vital statistics of
many countries worldwide from 1906 to 1936.]
Editions de l'Institut National d'Etudes D\'emographiques, Paris.

\qparr
Espinas (A.) 1878, 1935: Des soci\'et\'es animales [on animal
societies]. Thesis of the university of Paris. Republished
in 1935, F\'elix Alcan, Paris.\qL
[The title looks more original when one remembers that
the author was in fact a sociologist.]

\qparr
Herteliu (C.), Ileanu (B.V.), Ausloos (M.), Rotundo (G.) 2015:
Effect of religious rules on time of conception in Romania
from 1905 to 2001.
Human Reproduction 30,9,2202-2214.

\qparr
Houdaille (J.) 1979: Mouvement saisonnier des conceptions en France
de 1740 \`a 1829 [Seasonal pattern of conceptions in France from
1740 to 1829.]
Population 2,452-457.

\qparr
Houdaille (J.) 1985: Le mouvement saisonnier des naissances dans la
France rurale de 1640 \`a 1669. [Seasonal pattern of conceptions
in rural France from 1640 to 1669.]
Population 2,360-362.

\qparr
Huber (S.), Fieder (M.) 2011: Perinatal winter conditions after
later reproductive performance in Romanian women: intra and
intergenerational effects.
American Journal of Human Biology 23,546-552.

\qparr
Lam (D.A.), Miron (J.A.) 1994: Global patterns of seasonal
variation in human fertility.
Annals of the New York Academy of Sciences 709,9-28.

\qparr
Lerchl (A.) 2005: Where are the Sunday babies?
Observation on a marked decline on weekend births
in Germany.  
Naturwissenschaften 92,592-594.

\qparr
Lerchl (A.), Reinhard (S.C.) 2008: Where are the Sunday babies?
II. Declining weekend birth rates in Switzerland. 
Naturwissenschaften 95,161-164.

\qparr
Lerchl (A.) 2008: Where are the Sunday babies?
III. Caesarean sections, decreased weekend births, and midwife
involvement in Germany.
Naturwissenschaften 95,165-170.

\qparr
MacFarlane (W.V.) 1974: Seasonal cycles of human conception.
Progress in biometeorology A, Vol.1,557-577 and 711-713.

\qparr
R\'egnier-Loilier (A.) 2010a: \'Evolution de la saisonnalit\'e des
naissances en France de 1975 \`a nos jours [Changes in the seasonal
birth pattern in France from 1975 to 2006].
Population, 65,1,147-189.

\qparr
R\'egnier-Loilier (A.) 2010b: \'Evolution de la r\'epartition
des naissances dans l'ann\'ee en France
[Changes in the seasonal birth pattern in France].
Actes du XVe colloque national de d\'emographie [Proceedings
of the 15th national conference on demography, 24-26 May 2010]
Published by the ``Conf\'erence Universitaire de d\'emographie
et d'\'etude des populations''.

\qparr
Rey (G.), Fouillet (A.), Jougla (E.), H\'emon (D.) 2007:
Vagues de chaleur, fluctuations ordinaires des temp\'eratures
et mortalit\'e en France depuis 1971 [Heat waves, temperature
fluctuations and mortality in France from 1971 to 2003].
Population 62,3,533-564.

\qparr
Richmond (P.), Roehner (B.M.) 2018a:
Coupling between death spikes and birth troughs. 
Part 1: Evidence.
[Available on the arXiv website: arXiv: 1801.04533.]

\qparr
Richmond (P.), Roehner (B.M.) 2018b:
Coupling between death spikes and birth troughs. 
Part 2: Comparative analysis of salient features.
[Available on the arXiv website: arXiv: 1801.04535.]

\qparr
Sardon (J.-P.) 2005: Influence des \'epid\'emies de grippe
sur la f\'econdit\'e. [Influence of influenza epidemics on
fertility.]
In: Bergouignan (C.) et al.: ``La population de la France.
Evolutions d\'emographiques depuis 1946, Vol. 1, p. 413-417.\qL
[This short but interesting paper is focused
on the effect of the epidemic of 1957 on births in European
countries.]

\qparr
Seiver (D.A.) 1985: Trend and variation in the seasonality
of US fertility, 1947-1976.
Demography 22,1,89-100.

\qparr
Vicedo-Cabrera (A.), Ragettli (M.), Schindler (C.), R\"o\"osli (M.)
2016: Excess mortality during the warm summer of 2015 in
Switzerland.
Swiss Medical Weekly, 5 December 2016. 

\qparr
Villerm\'e (L.R.) 1931: De la distribution par mois des conceptions
et des naissances de l'homme [On the monthly distribution of
human conceptions and births].
Annales d'Hygi\`ene Publique et de M\'edecine L\'egale, S\'erie 1, No
5,55-155.

\end{document}